\documentclass[a4paper,11pt]{article}
\usepackage{pos}
\usepackage{hyperref}

\title{Usage of GPUs for ALICE Reconstruction in the ALICE Online Compute farm and in the Grid}

\author*[a]{David Rohr}
\onbehalf{the ALICE Collaboration}

\affiliation[a]{CERN\\
Esplanade des Particules 1, 1217 Meyrin, Switzerland}

\emailAdd{drohr@cern.ch}

\abstract{
ALICE, the dedicated heavy-ion experiment at the LHC, records Pb--Pb collisions at an interaction rate of up to 50 kHz.
It was the first LHC experiment to leverage GPUs for online data processing during LHC Runs 1 and 2, and its Run 3 online event reconstruction is now almost fully based on GPUs.
In order to use its online processing server farm also for offline processing in an efficient way while the LHC is not operating, ALICE has been running the offline TPC tracking on GPUs since 2023.
Since then, ALICE has been working on offloading more offline reconstruction steps to GPUs.
In particular, porting the track model decoding and ITS tracking to GPUs has improved the throughput by 29\%.
Starting in 2026, ALICE can use the GPUs at external grid sites beyond the ALICE online computing farm for offline reconstruction.   
A first production campaign ran on the NVIDIA GPUs of the NERSC Perlmutter cluster, marking the first use of GPUs for LHC offline reconstruction at an external grid site.
}

\FullConference{14th Edition of the Large Hadron Collider Physics (LHCP2026)\\
18-22 May 2026\\
Paris, France\\}

\begin{document}
\maketitle

\section{Introduction}

\looseness=-1
ALICE~\cite{bib:alice} is the dedicated heavy-ion experiment at the LHC.
ALICE upgraded both its detectors~\cite{bib:ls2upgrade} and the computing scheme~\cite{bib:o2tdr} to cope with much higher readout rates in Run 3.
Neither a hardware nor a software trigger is employed, but all recorded collisions are stored on the temporary disk buffer.
In the case of Pb--Pb, all collisions are stored permanently.
For pp, a skimming phase based on offline reconstruction selects which events to store permanently and which to drop.
The large data volumes and the absence of triggering make online compression the most important aspect of online computing.
By far the largest contributor to the ALICE data size is the TPC (Time Projection Chamber), and TPC processing is well suited to run on GPUs.
Therefore, ALICE operates a GPU-accelerated online processing farm called EPN~\cite{bib:epnfrontier} (Event Processing Nodes) equipped with 2800 AMD MI50 and MI100\footnote{https://www.amd.com/en/products/specifications/accelerators.html} GPUs and $\sim25000$ CPU cores.
The total raw readout data rate of ALICE is around 3.5 TB/s.
It is reduced to a maximum of 900 GB/s for high-rate Pb--Pb data by FPGAs via zero suppression, and afterwards the EPN farm compresses it to a maximum of 170~GB/s.
When there is no beam in the LHC, the EPN farm operates as a grid site for offline processing.

\section{GPU usage in offline processing at ALICE}

ALICE has a long history of GPU usage starting with the High Level Trigger~\cite{bib:hlt} of Run 1 and Run 2.
Since the beginning of Run 3, essentially all online processing runs on GPUs~\cite{bib:chep2023}.
With the EPN farm operating partially as a grid site and the majority of the compute performance of the EPNs provided by GPUs, it makes sense to use GPUs for offline reconstruction.
Since 2023, ALICE has been using the GPUs for TPC offline reconstruction in the EPNs~\cite{bib:chep2023}.
This constitutes between 50\% and 60\% (depending on the run conditions) of the total offline reconstruction workload.
Offloading it to GPUs efficiently removes this workload from the CPU, and increases the processing throughput in timeframes per second by 2x to 2.5x as demonstrated in EPN grid operation \cite{bib:chep2023}.

\looseness=-1
ALICE follows two strategies to improve GPU usage.
First, offloading more reconstruction steps beyond the TPC reconstruction to GPUs improves the GPU usage.
Second, with more grid sites offering GPUs today, it is a goal to also use GPUs at external grid sites and not only in the EPN farm.
For the first point, ALICE offloaded the TPC track model decoding~\cite{bib:gabrielelhcp}, a step in the decoding of compressed raw data, to GPUs in 2024, followed by ITS tracking~\cite{bib:chep2024its} in 2026.
Together, the reconstruction throughput, measured in timeframes processed per second, increases by 29\%.
For the use of GPUs at other grid sites, the workflow configuration must become more flexible, since the CPU-to-GPU performance ratio becomes a parameter and will differ from that on EPN nodes.

A first production of offline reconstruction ran in 2026 at the Perlmutter supercomputer\footnote{https://www.nersc.gov/what-we-do/resources/perlmutter} at NERSC.
The node configuration differs significantly from an EPN node.
While an EPN server has 2 CPU sockets with 2 times 4 AMD GPUs, the Perlmutter servers are single-socket with 4 NVIDIA GPUs.
For offline processing on the EPNs, each server is running two grid jobs in parallel, each using 64 virtual CPU cores (32 physical cores) and 4 GPUs.
Note that currently the EPN GPUs are not fully used in offline reconstruction.
Even with TPC + ITS tracking, only around 70\% of the workload runs on the GPU while the GPUs provide around 90\% of the compute capacity.
So the offline reconstruction on the EPNs is CPU-bound, and GPUs have significant idle times.
This will improve in the future, with more processing steps on the GPU.
The NVIDIA A100 GPUs at NERSC are faster than the EPN GPUs.
Based on initial benchmarks, one A100 GPU should be able to handle the full GPU workload.
Thus, the processing at NERSC currently uses 64 virtual CPU cores as on the EPNs, but only one GPU.
Obviously, this is not very efficient, since it leaves the majority of the GPU capacity at NERSC unused.
The only ways to improve this would be to significantly increase the fraction of work running on the GPU, or to run different very GPU-heavy jobs without much CPU load, possibly from other users, in parallel.

\begin{figure}[h]
\centering
\includegraphics[width=0.85\textwidth,clip]{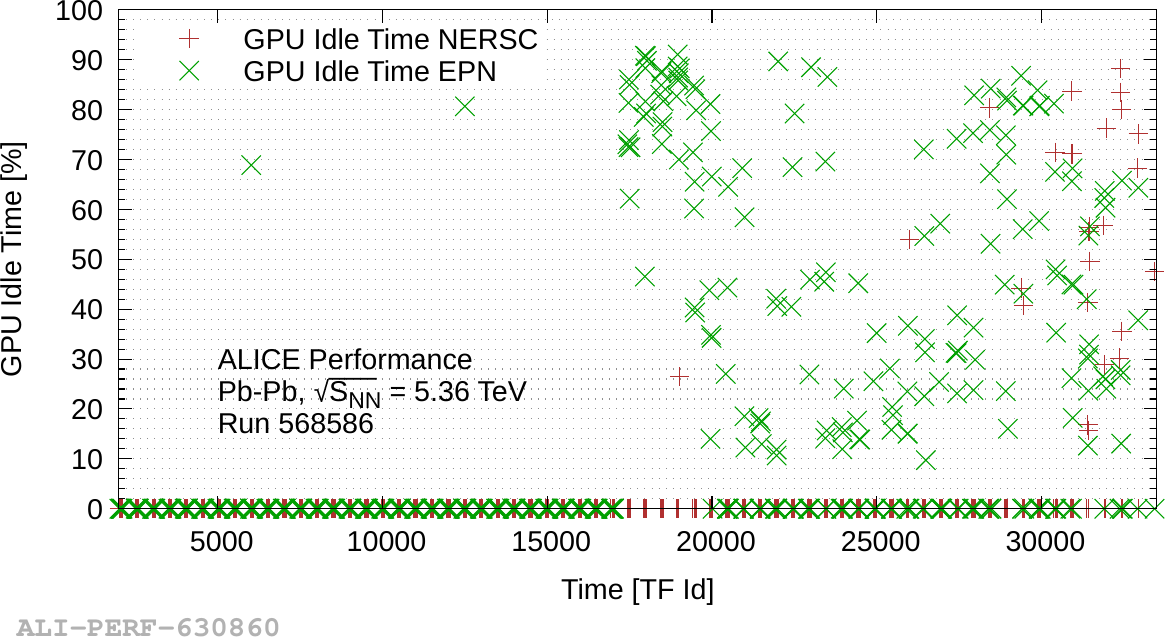}
\caption{GPU idle time of offline reconstruction jobs on EPN and at NERSC as a function of number of processed Time Frames (TF)}
\label{fig:gpu_idle_time}
\end{figure}

\looseness=-1
In the CPU-bound scenario, it is important that the GPU workload is offloaded efficiently to the GPU, without creating any bottlenecks for the remaining CPU tasks.
Figure~\ref{fig:gpu_idle_time} shows the GPU idle time during the first 35000 time frames at the NERSC site and the EPN site.
Note that the full production extends to the right beyond the shown range with similar behavior.
ALICE reconstruction uses a message-passing-based, pipelined, multi-process processing framework inside the ALICE Data Processing Layer (DPL)~\cite{bib:chep2023}.
This means different reconstruction tasks run as separate operating system processes, which exchange messages, in contrast to a single monolithic application where all data is kept in the same global memory.
The processes can run in parallel in a pipelined way, processing separate time frames in parallel, which exhibits an inherent parallelism.
The reconstruction chain is a directed graph, with time frames of 2.8 ms of consecutive data injected at the beginning of the graph.
Time frames are published by the source device of the processing chain, which loads them to memory.
Afterwards, they are in flight until they are fully processed, the sink device has written all output to disk, and all time frame data is removed from memory.
Rate limiting is implemented by defining a maximum number of time frames that are allowed to be in flight at the same time.
This is necessary to constrain the memory requirement and adapt it to the available memory in the server.
Once the time-frames-in-flight limit is reached, the rate limiting constraint prevents further publishing of time frames, and the source device will wait until the next time frame reaches the end of the chain.
When the pipelined processing starts, the first time frames are published in an unconstrained way, until the limit for time frames in flight is reached.
The TPC and ITS tracking on GPU are positioned at the beginning of the processing chain.
Thus, they process published time frames immediately, without waiting for any CPU task.
Therefore, at the beginning of the grid job, the GPUs are fully loaded.
Once the limit on the maximum number of time frames in flight is reached, time frame publishing gets throttled, and the GPUs begin to wait for input data.
This is visible as significant GPU idle time appearing once the time frame in flight limit is reached.
The job on the EPN using 4 GPUs has faster aggregate GPU processing than the job at NERSC with 1 GPU, thus it reaches the limit earlier, at around 17,000 time frames.
At NERSC, the limit is reached after 30,000 time frames.
From that time onward, the processing is fully CPU-bound on both sites with large GPU idle times, and the global throughput is fully determined by the CPU capacity.
To achieve full GPU utilization, it is important to tune the number of threads and the number of instances of slow reconstruction tasks, such that all tasks achieve the same aggregate throughput.
Otherwise, slow tasks become a bottleneck and backpressure the processing chain.
In a tuned setup, the average CPU load of the job (after some warm-up time and excluding the end when no new time frames are published and pipelines are flushed) is 90\% to 95\% (see Figure 2 of~\cite{bib:ichep2024}).

\section{ALICE3 and GPU usage in future experiments}

For LHC Run 4, ALICE will undergo minor updates, which do not impact the computing scheme.
For Run 5, the ALICE Collaboration is proposing to install a completely new experiment called ALICE 3~\cite{bib:alice3}.
ALICE 3 is an all-silicon pixel detector without a TPC, and it might necessitate a new computing scheme.
Depending on the noise level of the detector, which is not yet known, noise might contribute significantly to the raw data size.
While the ALICE 3 letter of intent assumed a similar compression scheme as today, the high noise level might require significant changes.
In addition, the current Run 3 compression and cluster rejection scheme~\cite{bib:ctd2019} might become infeasible.
It is based on the assumption that the online reconstruction finds at least a part of each track, such that all clusters in the vicinity of good tracks are protected.
However, in contrast to the TPC, finding tracks with a missing layer in a silicon detector increases the combinatorial complexity significantly.
In Run 3, ALICE does not allow for missing layers online.
However, offline tracking is much more time-consuming, and is thus most likely incompatible with online computing constraints.

\looseness=-1
A possible strategy could be a two-phase approach, like HLT1 and HLT2 in LHCb~\cite{bib:lhcb}, or similar to skimming for ALICE pp data taking in Run 3~\cite{bib:ppskimming}.
ALICE 3 would have a GPU-enabled processing farm at CERN for online and offline processing.
During data-taking, the online processing would do local raw data processing, clusterization, and simple compression before storing the intermediate data to a disk buffer.
Within a few days, the data would be reprocessed with an offline-quality tracking to allow for better compression and optionally cluster rejection.
Only afterwards the compressed raw data is written to permanent storage.
As in Run 3, offline processing would run partly at CERN leveraging GPUs, and partly in the grid.
The advantage would be that the online tracking does not need to cope with the peak LHC interaction rate, but only with the average interaction rate, averaged over beam and no-beam periods.
A negative side effect would be that if the online farm is dimensioned for the average rate, it would have much less spare capacity for offline GPU reconstruction.

\begin{figure}[h]
\centering
\includegraphics[width=0.84\textwidth,clip]{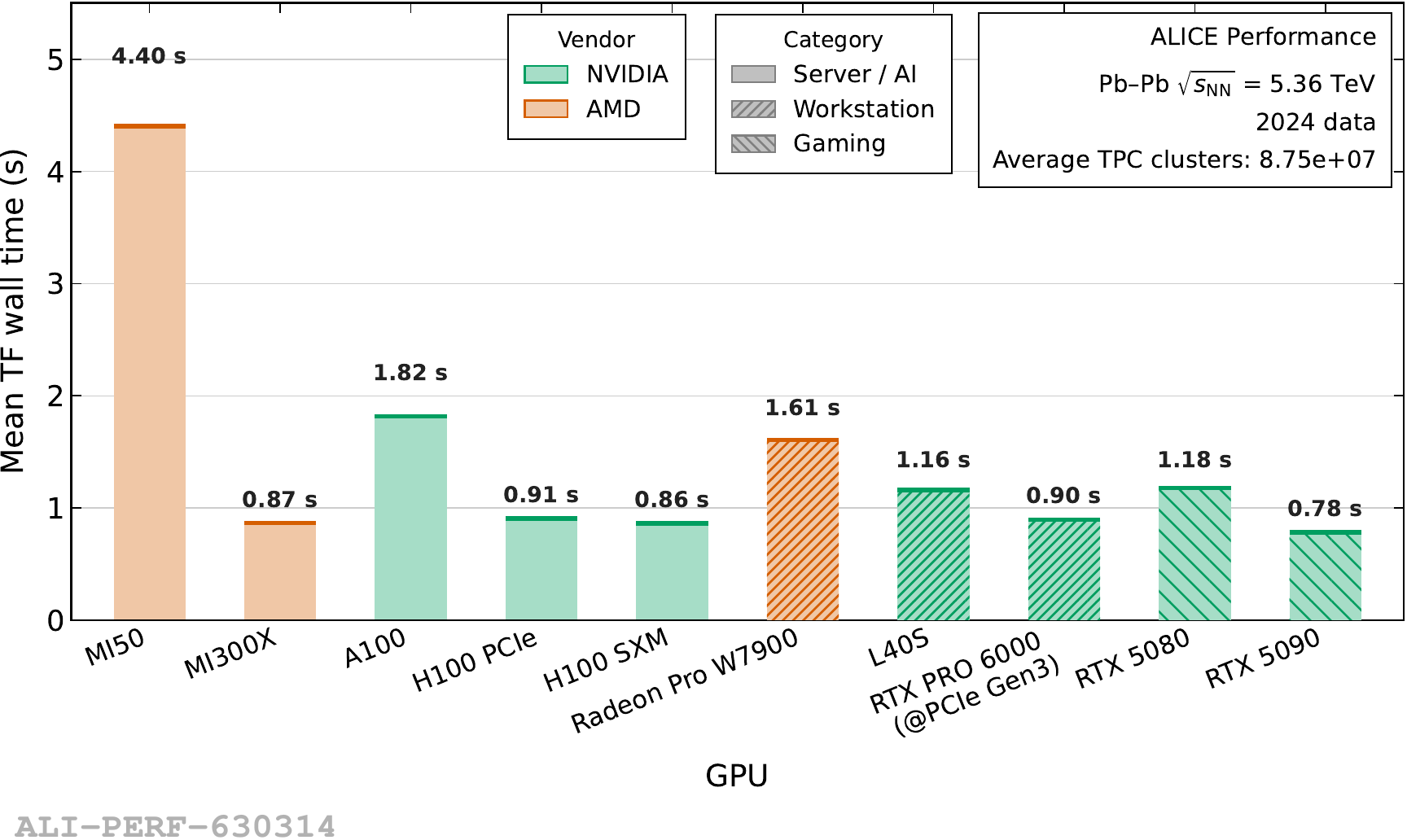}
\caption{Comparison of average processing time of a Pb--Pb time frame on different GPUs}
\label{fig:gpu_speed}
\end{figure}

\looseness=-1
ALICE benchmarks modern GPUs for different purposes: the foreseen upgrade of the current computing farm for Run 4, the computing farm of ALICE 3, and also for GPU usage in the grid.
Figure~\ref{fig:gpu_speed} gives an overview of different GPU models.
Today, the vendors focus more on AI and ML performance than on single-precision floating-point performance.
For ALICE, the server-grade AI GPU models such as H100 and MI300 do not provide a significant performance advantage over gaming GPUs to justify their much higher price.
In principle, gaming GPUs would be ideal, since professional GPU features such as ECC memory, large memory, and full-duplex PCIe transfers are not needed at all.
However, gaming GPUs do not come in passively cooled versions and are not qualified for operation in a data center.
The most promising solution for Kalman-filter based online track reconstruction seems to be server-grade workstation GPUs, which use the same GPU chips as the gaming GPUs, with more memory and passive cooling.
There are also use cases for machine learning, particularly for triggering, where GPUs optimized for AI and ML could play to their strengths.
Overall, single experiments are very small clients for the GPU vendors.
It might be helpful if the high energy physics community as a whole communicates to the GPU vendors the need for strong single-precision floating point in professional-grade GPUs for event reconstruction.
Otherwise, a few years from now, it might happen that there are no suitable GPU models available on the market any more.

\section{Conclusions}

\looseness=-1
ALICE uses GPUs heavily for online and offline processing in Run 3.
Currently, 99\% of the online reconstruction runs on GPUs, and ALICE is working to increase the GPU fraction of the offline reconstruction workload from currently 50\%-- 65\% to 80\%.
ALICE has been using the available spare GPU capacity of the online processing farm for offline reconstruction since 2023.
Starting in 2026, ALICE has also been able to use GPUs at external grid sites such as NERSC, and has run its first production campaign on the NVIDIA GPUs of the Perlmutter HPC site.
ALICE is investigating which GPUs are most promising for the future in Run 4 and for ALICE 3.
Since the main requirement is single-precision floating point performance, the most promising options at the moment would be server-grade workstation GPUs, employing the same GPU chips as the gaming GPUs.

\end{document}